\documentstyle[11pt,aaspp4,flushrt]{article}  % preprint format, one column
\journalid{VOL}{JOURNAL DATE 1997}
\articleid{START PAGE}{END PAGE}
\font\fiverm=cmr5

\def\pr{Phys. Rev.}                             % DO NOT DELETE
\def\aapl{Astron. Astr. (Lett.)}                % DO NOT DELETE
\def\ssr{Space Sci. Rev.}                       % DO NOT DELETE
\def\teq#1{$\, #1\,$}                           % text equation
\input psfig.tex                           % for figure embedding
{\catcode`\@=11                                                  
\gdef\SchlangeUnter#1#2{\lower2pt\vbox{\baselineskip 0pt\lineskip0pt    
\ialign{$\m@th#1\hfil##\hfil$\crcr#2\crcr\sim\crcr}}}}           
\def\gtrsim{\mathrel{\mathpalette\SchlangeUnter>}}

\def\erg{\varepsilon}

\def\dover#1#2{\hbox{${{\displaystyle#1 \vphantom{(} }\over{
   \displaystyle #2 \vphantom{(} }}$}}                
\def\sigt{\hbox{$\sigma_{\hbox{\fiverm T}}$}}                                   
\def\taupp{\tau_{\gamma\gamma}}
\def\sigpp{\sigma_{\gamma\gamma}}

\def\eb{\erg_{\hbox{\fiverm B}}}
\def\Eb{E_{\hbox{\fiverm B}}}
\def\ec{\erg_{\hbox{\fiverm C}}}
\def\Ec{E_{\hbox{\fiverm C}}}
\def\etab{\eta_{\hbox{\fiverm B}}}
\def\etac{\eta_{\hbox{\fiverm C}}}
\begin{document}
%
%      WARNING: THE FOLLOWING MACROS ARE ESSENTIAL FOR AASTEX
%            INTERPRETATION OF THIS DOCUMENT:    DO NOT DELETE!!
%
\newcommand{\vol}[2]{$\,$\rm #1\rm , #2.}                 
\newcommand{\figureout}[2]{  \begin{figure}  \epsfysize=12.5cm
   \hbox to\hsize{\hfill \hbox{\epsfbox{#1}} \hfill} 
   \vskip 0.0cm \caption{#2} \end{figure} \clearpage }
\newcommand{\tableout}[4]{\vskip 0.3truecm \centerline{\rm TABLE #1\rm}
   \vskip 0.2truecm\centerline{\rm #2\rm}   %\baselineskip=\tablineskip
   \vskip -0.3truecm  \begin{displaymath} #3 \end{displaymath} 
   \noindent \rm #4\rm\vskip 0.1truecm } %  \baselineskip=\textlineskip} 
%
%      END OF ESSENTIAL MACROS.
%
\title{PAIR PRODUCTION ABSORPTION TROUGHS IN GAMMA-RAY BURST\\
   SPECTRA: A POTENTIAL DISTANCE DISCRIMINATOR}
   \author{Matthew G. Baring\altaffilmark{1} and Alice K. Harding}
   \affil{Laboratory for High Energy Astrophysics, Code 661, \\
      NASA Goddard Space Flight Center, Greenbelt, MD 20771, U.S.A.\\
      \it baring@lheavx.gsfc.nasa.gov, harding@twinkie.gsfc.nasa.gov\rm}
   \altaffiltext{1}{Compton Fellow, Universities Space Research Association}
   \authoraddr{Laboratory for High Energy Astrophysics, Code 661,
      NASA Goddard Space Flight Center, Greenbelt, MD 20771, U.S.A.}
% 
%\slugcomment{To appear in \it The Astrophysical Journal\rm , *** issue.}
%\slugcomment{Accepted for publication in \it The Astrophysical Journal\rm}
%
%\clearpage

\begin{abstract}  
In order to explain the emergence of a high-energy continuum in
gamma-ray bursts (GRBs) detected by EGRET, relativistic bulk motion
with large Lorentz factors has recently been inferred for these sources
regardless of whether they are of galactic or cosmological origin.
This conclusion results from calculations of internal pair production
opacities in bursts that usually assume an infinite power-law source
spectrum for simplicity, an approximation that is quite adequate for
some bursts detected by EGRET.  However, for a given bulk Lorentz
factor \teq{\Gamma}, photons above the EGRET range can potentially
interact with sub-MeV photons in such opacity calculations.  Hence it
is essential to accurately address the spectral curvature in bursts
seen by BATSE, and also treat the X-ray paucity that is inferred from
low energy fluxes observed in the X-ray band.  In this paper we present
the major properties induced in photon-photon opacity considerations by
such spectral curvature.  The observed spectral breaks around 1 MeV
turn out to be irrelevant to opacity in cosmological bursts, but are
crucial to estimates of source transparency in the 1 GeV -- 1 TeV range
for sources located in the galactic halo.  We find that broad
absorption troughs can arise at these energies for suitable bulk motion
parameters \teq{\Gamma}.  Such troughs are probably an unambiguous
signature of a galactic halo population, and if observed by experiments
such as Whipple, MILAGRO and GLAST, would provide powerful evidence
that such bursts are not at cosmological distances.
\end{abstract}  
\keywords{gamma-rays: bursts --- radiation mechanisms: non-thermal --- 
gamma rays: theory --- relativity}
\clearpage 
\section{INTRODUCTION}

High energy gamma-rays have been observed for six gamma-ray burst
sources by the EGRET experiment on board the Compton Gamma-Ray
Observatory (CGRO).  Most conspicuous among these observations is the
emission of an 18 GeV photon by the GRB940217 burst (Hurley et al.
1994).  These detections occurred during the first five years of the
mission, when the EGRET spark chamber gas level was not severely
depleted and, taking into account the instrumental field of view, they
indicate that emission in the 1 MeV--10 GeV range is probably common
among bursts, if not universal.  One implication of GRB observability
at energies around or above 1 MeV is that, at these energies,
two-photon pair production (\teq{\gamma\gamma\to e^+e^-}) is not
producing any significant spectral attenuation in the source.
Attenuation by pair creation in the context of GRBs was first explored
by Schmidt (1978).  He assumed that a typical burst produced
quasi-isotropic radiation, and concluded at the time that the detection
of photons around 1 MeV limited bursts to distances less than a few kpc,
since the optical depth scales as the square of the distance to the burst.

The observation by EGRET of emission above 100 MeV clearly indicated
that Schmidt's analysis needed serious revision.  Furthermore, BATSE's
determination of the spatial isotropy and inhomogeneity of bursts (e.g.
Meegan et al. 1991, 1996) suggested that they are either in an
extended halo or at cosmological distances.  Consequently their
intrinsic luminosities, and therefore their optical depths to pair
production for isotropic radiation fields, are much higher than
previously believed.  Hence the suggestion that GRB photon angular
distributions were highly beamed and produced by a relativistically
moving or expanding plasma (e.g.  Fenimore et al. 1992) has become very
popular.  Various determinations of the bulk Lorentz factor
\teq{\Gamma} of the medium supporting the GRB radiation field have been
made in recent years, mostly concentrating (e.g. Krolik and Pier 1991,
Baring 1993, Baring and Harding 1993, Harding 1994) on the simplest
case where the angular extent of the source was of the order of
\teq{1/\Gamma}.  These calculations assumed an infinite power-law burst
spectrum, and deduced that gamma-ray transparency up to the maximum
energy detected by EGRET required \teq{\Gamma\gtrsim 5} for galactic
halo sources and \teq{\Gamma\gtrsim 200} for cosmological bursts.
These limits are reproduced by a wide range of source geometries
(Baring and Harding 1996).

While the assumption of an infinite power-law source spectrum is
expedient, the spectral curvature seen in most GRBs by BATSE (e.g. Band
et al. 1993) is expected to play an important role in reducing
estimates for \teq{\Gamma} for potential TeV emission from these
sources.  Such curvature is patently evident in 200 keV--2 MeV spectra
of EGRET-detected bursts, and its prevalence in bursts is indicated by
the generally steep EGRET spectra for bursts (e.g. Schneid et al. 1992;
Kwok et al. 1993; Sommer et al. 1994; Hurley et al. 1994).
Furthermore, the relative paucity of emission detected below 10 keV in
a number of GRBs (e.g. Epstein 1986) can potentially lower the opacity
for the highest energy photons substantially.  In this paper, the
principal effects introduced into pair production opacity calculations
by spectral curvature over the BATSE energy range and the lower
portions of the EGRET domain are considered.  We find that the presence
of such curvature generally has minimal influence on the spectra and
inferred bulk motions for bursts of cosmological origin.  In contrast,
for galactic halo sources, we observe that for realistic parameters of
the bulk motion, source opacity may arise only in a portion of the 1
GeV -- 1 TeV range, with transparency returning in the super-TeV range,
resulting in the appearance of distinctive broad absorption troughs.
Such features may provide a unique identifier for bursts in halo
locales, so that current and future ground-based initiatives such as
Whipple and MILAGRO, and space missions such as GLAST may play a key
role in determining the distance scale for gamma-ray bursts.

\section{$\gamma$-$\gamma$ ATTENUATION AND SPECTRAL CURVATURE}
                                                 
In assessing the role of two-photon pair production in burst spectral
attenuation, the interactions of photons created only within the
emission region are considered here.  Recent authors have invoked
relativistic beaming in sources when superseding Schmidt's (1978) early
work.  This hypothesis builds on the property that \teq{\gamma\gamma\to
e^-e^+} has a threshold energy \teq{E_1} that is strongly dependent on
the angle \teq{\Theta} between the photon directions:  \teq{E_1 >
2m_e^2c^4/[1-\cos\Theta] E_2} for target photons of energy \teq{E_2}.
Hence radiation beaming associated with relativistic bulk motion of the
underlying medium can dramatically reduce the optical depth
\teq{\taupp} in sources at enormous distances from earth, suppressing
$\gamma$-ray spectral attenuation turnovers and blueshifting them out
of the observed spectral range.  The simplest picture of relativistic
beaming has ``blobs'' of material moving with a bulk Lorentz factor
\teq{\Gamma} more-or-less toward the observer, and having an angular
``extent'' \teq{\sim 1/\Gamma} (Krolik and Pier 1991, Baring 1993,
Baring and Harding 1993).  These works assumed an infinite power-law
spectrum \teq{n(\erg )=n_{\gamma} \,\erg^{-\alpha }}, where \teq{\erg}
is the photon energy in units of \teq{m_ec^2}, for which the optical
depth to pair creation assumes the form \teq{\taupp (\erg )\propto
\erg^{\alpha -1}\Gamma^{-(1+2\alpha )}} for \teq{\Gamma\gg 1}.

Approximating the source photon spectrum by an infinite power-law is
expedient, however most bursts detected by BATSE show significant
spectral curvature in the 30 keV--500 keV range (e.g. Band et al.
1993).  Furthermore, BATSE sees MeV-type (i.e. 500 keV--2 MeV) spectral
curvature with significant frequency in bright bursts: see Schaefer et
al. (1992) for an analysis of a brightness-selected sample from early
in the BATSE era.  EGRET observes three of the Schaefer et al. sources
with such ``high energy'' breaks (Schneid et al. 1992; Kwok et al.
1992), and later EGRET bursts also exhibit ``MeV-type breaks'' ---
for relevant parameters, see Table~1.  Hence, EGRET detections seem
correlated with spectral breaks at the upper end of the BATSE energy
range, which is probably a selection effect for the observability of
bursts by EGRET: GRBs with breaks at higher energies tend to be more
luminous in the super-MeV range.  Whether bursts with breaks at MeV
energies are a class of objects distinct from the majority that
turnover at lower energies remains to be seen.  The MeV-type breaks and
those generally seen at lower energies in the BATSE data for many GRB
spectra could, in principle, reduce the opacity of potential TeV
emission from these sources, so it is important to generalize the pair
production opacity/relativistic beaming analysis to include the effects
of spectral curvature. 

The effects of a depletion of low energy photons in the BATSE range
relative to the EGRET quasi-power-law spectra can quickly be determined
by taking the simplest approximation to spectral curvature, namely a
power-law broken at a dimensionless energy \teq{\eb} (\teq{=\Eb /0.511}MeV)
with a low energy cut-off at \teq{\ec}:
\begin{equation}
   n(\erg )\; =\; n_{\gamma}\eb^{-\alpha_h}\, \cases{
   0, & if \teq{\erg\leq\ec\;\;},\cr
   \eb^{\alpha_l}\erg^{-\alpha_l} , &
     if \teq{\vphantom{\Bigl(} \ec\leq\erg\leq\eb\;\;},\cr
   \eb^{\alpha_h}\erg^{-\alpha_h} , & \teq{\erg >\eb\;\; }. \cr }
 \label{eq:powerlaw}
\end{equation}
The optical depth determination for such a distribution follows the
above description for pure power-laws, and utilizes results obtained in
Gould and Schreder (1967) and Baring (1994) for truncated power-laws.
The details of our calculations are presented in Baring and Harding
(1997, in preparation, hereafter BH97), where the optical depth
\teq{\taupp (\erg )} for pair production attenuation of a broken
power-law photon distribution is found to be
\begin{equation}
   \dover{\taupp (\erg )}{n_{\gamma}\sigt R}\;\approx \;
   \dover{{\cal A}(\alpha_l)}{\eb^{\alpha_h-\alpha_l}}\, 
\Bigl\{ {\cal H}(\alpha_l,\, \etac)
   - {\cal H}(\alpha_l,\, \etab )\Bigl\}\, 
   \dover{\erg^{\alpha_l-1}}{\Gamma^{2\alpha_l}}
   + {\cal A}(\alpha_h)\, {\cal H}(\alpha_h,\, \etab
   )\, \dover{\erg^{\alpha_h-1}}{\Gamma^{2\alpha_h}} \quad ,
 \label{eq:taupp}
\end{equation}
where \teq{\etab = \max\{1,\; \sqrt{\eb\erg}\, /\Gamma\,\} } and
\teq{\etac = \max\{1,\; \sqrt{\ec\erg}\, /\Gamma\,\} }, and
\begin{equation}                                           
   {\cal H}(\alpha ,\,\eta )\; =\;\dover{4}{\sigt}\int_1^{\infty} d\chi\,
   \dover{q^{2(1+\alpha )}}{\chi^{2\alpha-1}}\,\sigpp (\chi )\; ,\quad
   q = \min\Bigl(1,\; \dover{\chi}{\eta}\Bigr) \; ,
 \label{eq:Hfunc}
\end{equation}
for \teq{\sigpp} being the pair production cross-section.  Here
\teq{\chi =[\erg\omega (1-\cos\Theta)/2]^{1/2}} is the
center-of-momentum frame (CM) energy of the photons, for \teq{\Theta}
being the angle between the directions of photons of dimensionless
energies \teq{\erg} (test photon) and \teq{\omega} (interacting).  The
pair production threshold condition is then \teq{\chi\geq 1}.  The
function \teq{{\cal H}(\alpha ,\,\eta )} is equivalent, up to a factor
of 4, to that in Eq.~(23) of Gould and Schreder (1967). The
specialization \teq{{\cal H}(\alpha ,\, 1)} can be approximated by
\teq{7/(6\alpha^{5/3})} (e.g. Baring 1994) to better than 1\% for
\teq{1<\alpha <7}.  Setting \teq{\alpha_l=\alpha_h} also reproduces the
infinite power-law dependence \teq{\taupp(\erg )\propto
\erg^{\alpha_h-1}\Gamma^{-1-2\alpha_h}}, remembering that
\teq{n_{\gamma}\propto fd^2/R^2} for fluxes measured at earth
of\teq{f=4\pi n_{\gamma}cR^2/d^2} at 511 keV per 511 keV and source
size \teq{R=R_v\Gamma} and source distance \teq{d}.  We choose a
variability ``size'' \teq{R_v=3\times 10^7}cm here, following Baring
and Harding (1996).  Note that \teq{{\cal A}(\alpha )} is a function
that arises from the integrations over photon angles, and is
approximately given (to 1\%, Baring 1994) by \teq{2/(4/3+\alpha
)^{27/11}}.  More gradual spectral curvature can be treated by fitting
the GRB continuum with piecewise continuous power-laws, generalizing
the structure inherent in Eq.~(\ref{eq:taupp}); our technique is
well-suited to this adaptation (discussed in BH97).

A remarkable feature of Eq.~(\ref{eq:taupp}) is that the optical depth
is no longer necessarily a monotonically increasing function of energy
\teq{\erg}.  The parameters \teq{\etab} and \teq{\etac} govern the
importance, or otherwise, of spectral curvature effects.  Clearly, if
either \teq{\eb} or \teq{\erg} (in \teq{m_ec^2}) is low enough or
\teq{\Gamma} large enough that \teq{\etab =1}, then the pure power-law
result emerges from Eq.~(\ref{eq:taupp}) and curvature effects are
negligible.  Such a situation might then be expected for cosmological
bursts.  Conversely if \teq{\etab} or \teq{\etac} exceed unity, the
shape of the BATSE spectrum becomes crucial to optical depth
estimates.  The spectra in Eq.~(\ref{eq:powerlaw}) were attenuated
using the optical depth in Eq.~(\ref{eq:taupp}), specifically via an
exponential factor \teq{\exp\{-\taupp\} }, for fluxes and spectra
typical of bright gamma-ray bursts (e.g. see Table 1); the emergent
spectra are depicted in Fig.~\ref{fig:attengeneral}.  The results for
the cosmological case are noticeably uninteresting: the input broken
power-law is modified (as expected intuitively) by a quasi-exponential
turnover at an energy that is an increasing function of the bulk
Lorentz factor \teq{\Gamma} of the expansion that generates the
radiation.  If \teq{\Gamma} were chosen large enough to permit emission
out to TeV energies, the spectra would suffer attenuation due to the
\it external \rm supply of cosmological infra-red background photons
(Stecker and De Jager 1996, Mannheim et al. 1996).  Note that opacity
skin effects can sometimes render the exponential \teq{\exp\{-\taupp\}
} a poor descriptor of attenuation, with \teq{1/(1+\taupp )} perhaps
being an improvement, leading to broken power-laws rather than
exponential turnovers.  Such alternatives, which are model-dependent,
do not qualitatively affect the conclusions drawn here and are
discussed in BH97.
  
\placefigure{fig:attengeneral}

The galactic halo case in Fig.~\ref{fig:attengeneral}, where typically
\teq{\etab\gtrsim 1}, exhibits remarkably different behaviour: broad
absorption features occur in the 1 GeV--1 TeV range, depending on the
choice of source \teq{\Gamma}.  The presence of these notable troughs,
which become more pronounced as \teq{\Gamma} decreases, results from
the non-monotonic behaviour of the optical depth with energy:
\teq{\taupp} drops below unity around the TeV range due to the
``depleted supply'' of interacting photons in the low energy BATSE
portion of the spectrum.  These distinctive features arise only for
large spectral breaks (i.e.  \teq{\delta\alpha =\vert
\alpha_h-\alpha_l\vert\gtrsim 1.3}), and also MeV-type break energies;
a reduction of the severity or energy of the break pushes the
attenuation towards the more familiar exponential turnovers depicted in
Fig.~\ref{fig:attengeneral}.  The appearance of the troughs is most
strongly dependent on the size of \teq{\delta\alpha} and on
\teq{\alpha_l} being not too large.  Consequently, from the data in
Table 1, it appears that GRBs 910601 and 910814 would be the strongest
candidates for producing features like those exhibited in
Fig.~\ref{fig:attengeneral}. The other EGRET bursts are in a marginal
regime of parameter space; attenuation results for them are discussed
in BH97.  Note that the structure on the low energy end of the troughs
is a product of the sharp spectral breaks used, and is smoothed out
(BH97) for more realistic spectral curvature.

\placefigure{fig:atten910814}

Attenuation of spectra appropriate to the burst GRB 910814 are depicted
in Fig.~\ref{fig:atten910814} for different \teq{\Gamma}.  The case
where there is no low energy cutoff yields an extremely broad trough,
that is more reminiscent of a ``shelf.''  Such behaviour differs from
the halo cases in Fig.~\ref{fig:attengeneral} because here the source
spectra are somewhat steeper, producing much broader troughs; the
result would be an improbability of observing sources like GRB 910814
at TeV energies when no low energy cutoff is present.  In contrast,
when a spectral cutoff is present at 5 keV in
Fig.~\ref{fig:atten910814} (which is below the threshold of BATSE
sensitivity), mimicking the X-ray paucity of bursts (Epstein 1986;
actually this is a paucity of energy in X-rays, so that it may be
better described by a significant spectral break), the
supply of interacting photons is further depleted, the troughs narrow
and TeV emission reappears.  This property clearly underlines the
importance of the relationship between the 10 GeV--1 TeV and the soft
X-ray portions of the GRB spectrum, correlations that can only be
analyzed using broad band observations of bursts.  Note that the
highest EGRET spectral point for GRB 910814 in
Fig.~\ref{fig:atten910814} indicates that the bulk Lorentz factor is
constrained to \teq{\Gamma\gtrsim 20}.

The array of spectral shapes depicted in Figs.~1 and~2 indicates that a
diversity of such forms must be anticipated in GRBs, with troughs,
shelfs and turnovers, depending on source spectral
parameters (see BH97).  The reason for the appearance of troughs in
the galactic halo case but not for cosmological distances stems from
the much higher photon densities inferred in cosmological sources for a
given flux at earth, for which photon-photon collisions primarily
involve photons well above 1 MeV, and spectral curvature in the BATSE
range is irrelevant.  Note that the model-dependent details of pair
cascading have been neglected here; these are discussed in BH97.

\section{IMPLICATIONS}

The importance of the spectral attenuation results presented here is
immediately apparent.  The absorption troughs in Figs.~1 and~2 cannot
be produced for large source distances and are unambiguous markers of a
galactic halo burst population; they are consequently a potentially
powerful observational diagnostic.  Observations by the air
\v{Cerenkov} detectors Whipple and MILAGRO (a new water tank
experiment) at the 300 GeV--5 TeV range, and perhaps even HEGRA at
higher energies, combined with a probing of the 100 MeV -- 100 GeV
range by future space instrumentation such as GLAST (a spark chamber
experiment) could confirm or deny the existence of such absorption
features.  Note that the observation of apparently sharp cutoffs would
not distinguish between cosmological or galactic burst hypotheses.  The
current Whipple sensitivity threshold (Connaughton et al. 1995) is not
sufficiently constraining, so future generations of experimentation
will be required to assess whether or not GeV--TeV troughs are present
in GRB spectra.  Having five times the field of view and around ten
times the sensitivity of EGRET, and therefore the potential to detect
dozens of bright bursts per year, the proposed GLAST mission (Michelson
1996) might be expected to make significant inroads into this problem.

The shape and position of these prominent features are strongly
dependent on the spectral slopes and fluxes in the BATSE and EGRET
ranges, and even more interestingly, on the contributions to the source
flux from below 10 keV.  Hence coupled X-ray, soft and hard gamma-ray
observations are clearly warranted, an exciting challenge to the
astronomy and GRB communities.  The strong and well-defined
correlations between the width and shape of the troughs and the
spectral curvature below 10 MeV act as clear markers of the pair
production attenuation analysis discussed here, and are unlikely to be
replicated in detail by alternative, multi-component models (e.g. Katz
1994, M\'esz\'aros and Rees 1994) of hard gamma-ray emission, or from
opacity due to external background fields of radiation (such
attenuation is generally above 100 GeV: Stecker and De Jager 1996,
Mannheim, et al. 1996).  In summary, this paper presents a patently
powerful spectral diagnostic in the super-GeV range that can provide
enticing prospects for solving the problem of gamma-ray burst location.

\acknowledgments
We thank Dieter Hartmann, Bob Gould, Jay Norris and David Thompson for
reading the manuscript, and Brenda Dingus and Jennifer Catelli for
discussions about EGRET burst data.  

\newpage
\baselineskip 14pt plus0.3pt minus0.3pt

\clearpage

% Table 1

\psfig{figure=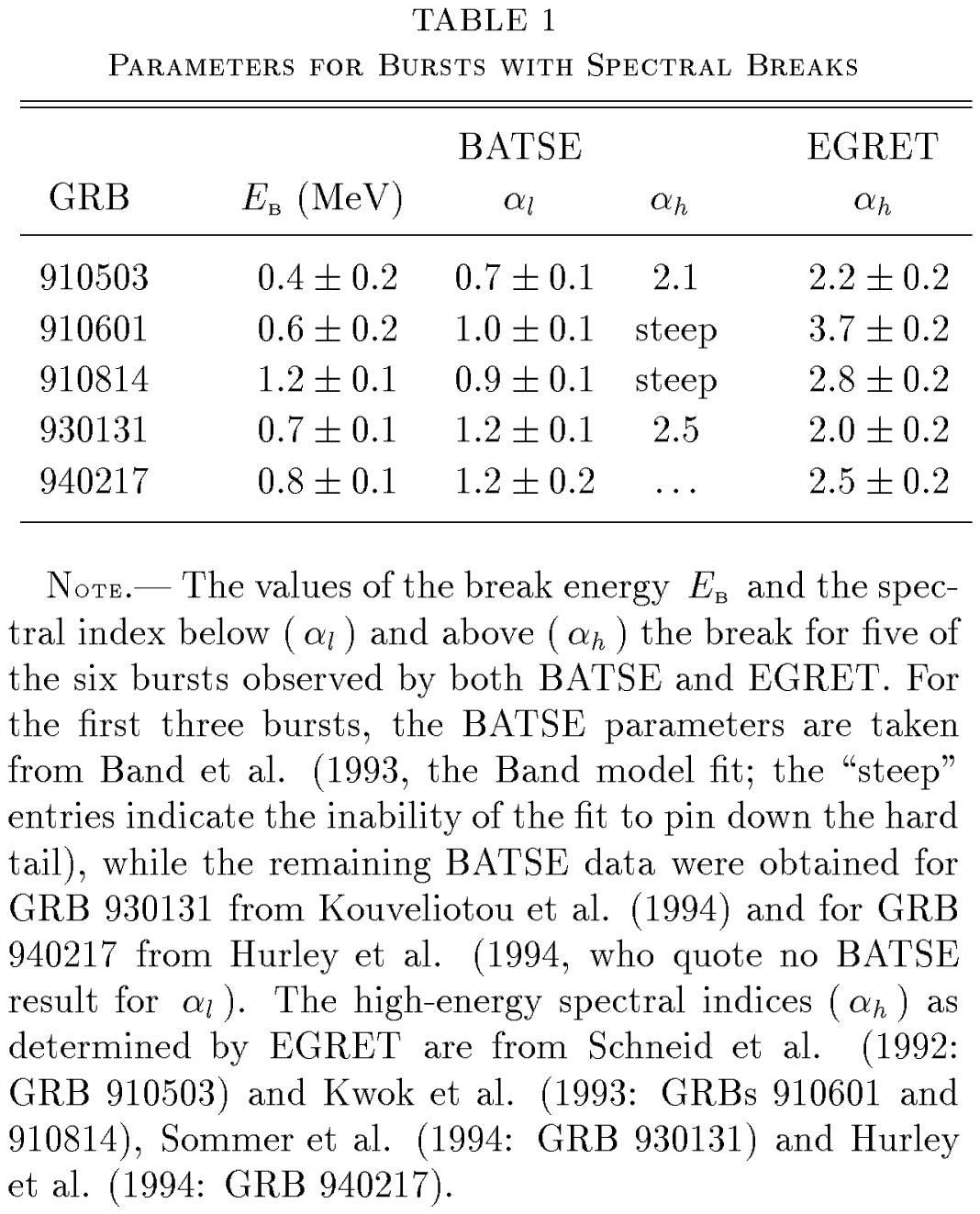,width=16.4truecm}

\clearpage

% Figures

\figureout{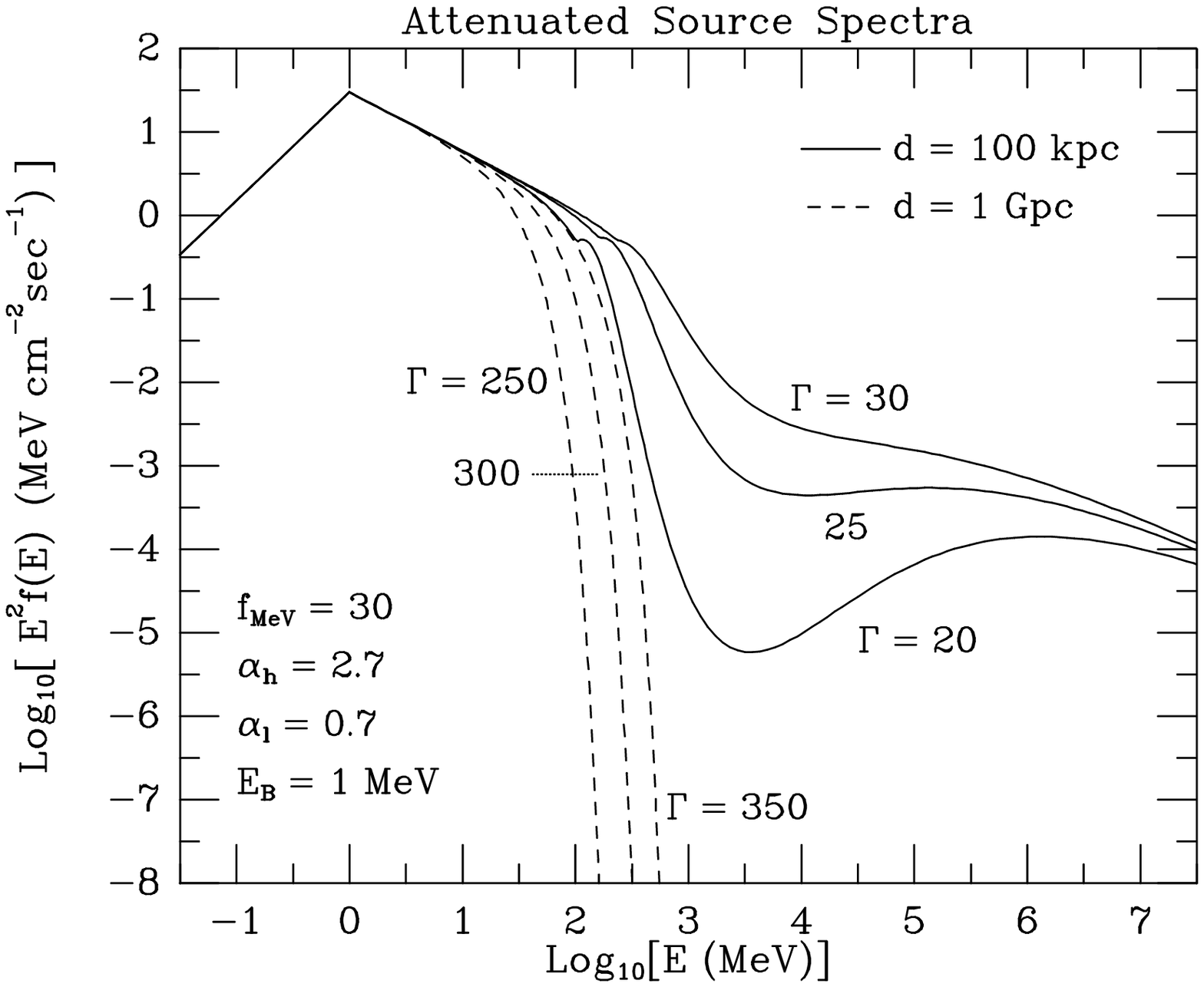}{
           The attenuation, internal to the source, of a broken power-law 
spectrum [see Eq.~(\ref{eq:powerlaw}) for definition], for source
distances typical of galactic halo (solid curves) and cosmological
(dashed curves) populations, and different bulk Lorentz factors
\teq{\Gamma} for the emitting region, as labelled.  The spectra,
plotted in the \teq{E^2\, f(E)} (i.e. \teq{\nu F_{\nu}}) format, are
attenuated by pair creation according to the factor \teq{\exp\{
-\taupp\} } for optical depths determined via Eq.~(\ref{eq:taupp})
(i.e. no pair cascading is included).  Here no low energy cutoff was
used, i.e. \teq{\ec =0}.  The quasi-exponential turnovers of the
cosmological cases can provide lower bounds to \teq{\Gamma} using
current EGRET data, and contrast the broad absorption troughs of the
halo examples.
    \label{fig:attengeneral}
}

\figureout{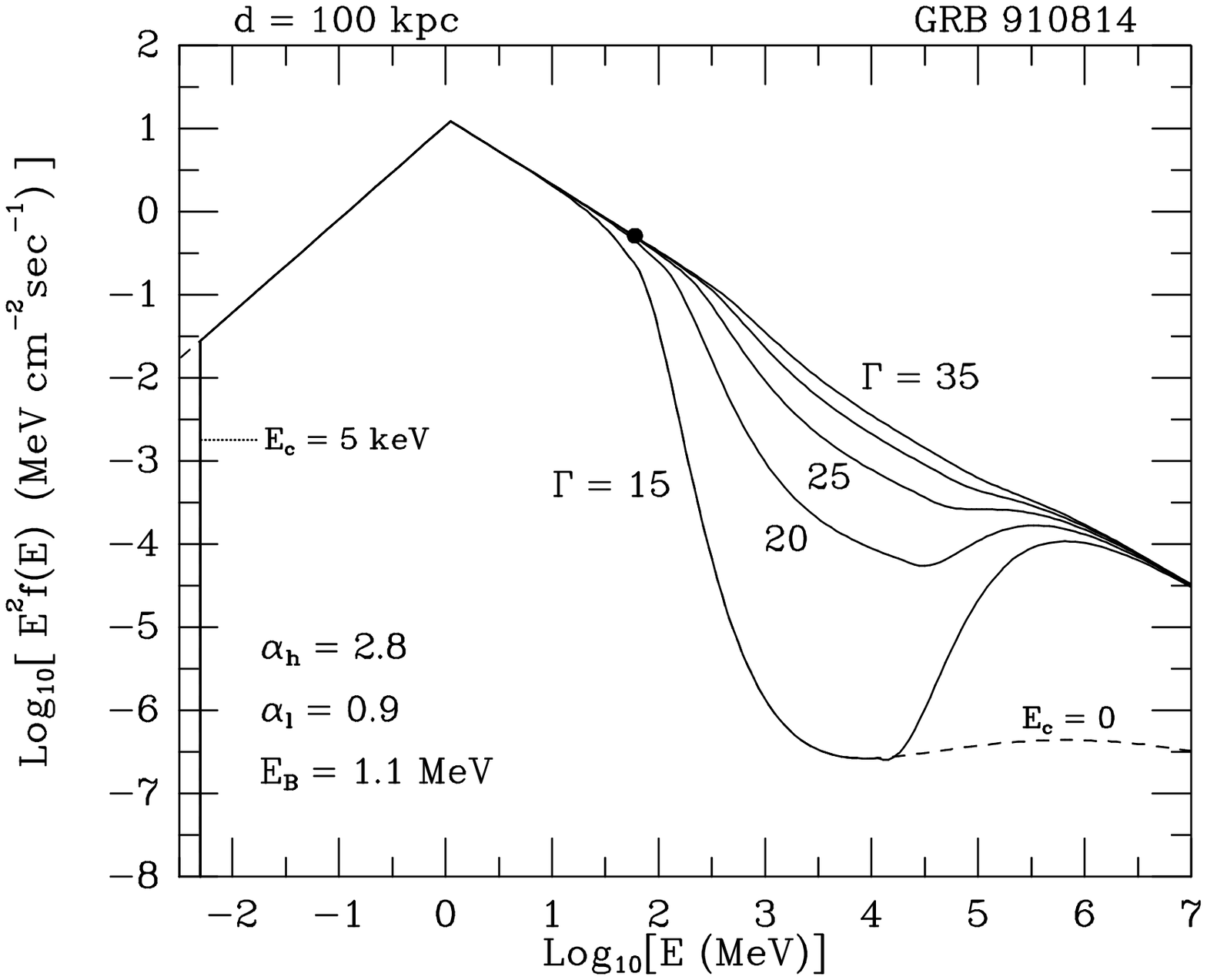}{
          Pair production (internal) attenuation of a broken power-law
fit (again in \teq{E^2\, f(E)} format) to the GRB 910814 spectrum, for
a galactic halo source distance of 100 kpc, for radiation emanating
from regions of different bulk Lorentz factors \teq{\Gamma},
incremented by 5, as labelled.  The power-law parameters for
Eq.~(\ref{eq:powerlaw}) were obtained from Table~1 and the solid curves
correspond to a low energy cutoff at \teq{\Ec =5}keV.  The dashed curve
displays the \teq{\Gamma =15} case for no cutoff (i.e. \teq{\Ec =0}).
The dot marks the highest EGRET data point, at 60 MeV.
    \label{fig:atten910814}
}

\end{document}